# Channelling chaos by building barriers

C. Chandre*, G. Ciraolo*, F. Doveil‡, R. Lima*, A. Macor‡ & M. Vittot*

\* *Centre de Physique Théorique-CNRS, Luminy-case 907, F-13288 Marseille cedex 9, France*

‡ *Physique des Interactions Ioniques et Moléculaires, Unité 6633 CNRS-Université de Provence, Equipe turbulence plasma, case 321, Centre Saint-Jérôme, F-13397 Marseille cedex 20, France*

Abstract

Chaos often represents a severe obstacle for the set-up of many-body experiments, e.g., in fusion plasmas or turbulent flows. We propose a strategy to control chaotic diffusion in conservative systems. The core of our approach is a small apt modification of the system which channels chaos by building barriers to diffusion. It leads to practical prescriptions for an experimental apparatus to operate in a regular regime (drastic enhancement of confinement). The experimental realization of this control on a Travelling Wave Tube opens the possibility to practically achieve the control of a wide range of systems at a low additional cost of energy.

The sensitivity of chaotic systems to small perturbations triggered a strong interdisciplinary effort to control chaos [1-8]. After the seminal work on optimal control by Pontryagin [9], new and efficient methods were proposed for controlling chaotic systems by nudging targeted trajectories [10-13]. However, for many body experiments like the magnetic confinement of a plasma [14] or the control of turbulent flows, such methods are hopeless due to the high number of trajectories to deal with simultaneously. In this Letter, we show a general strategy for conservative systems: Instead of tilting a trajectory, we build barriers in phase space by adding a small apt perturbation, hence confining all the trajectories. An experimental check of our method is performed on a paradigm for the transition to chaos [15]: a beam of test electrons interacting with electrostatic waves where chaos destroys the kinetic coherence of the electrons. By adding a small controlling wave the kinetic coherence of the beam is then restored.

The core of our strategy of control is to build barriers bounding the motion in phase space, at least for exponentially large times (the so-called Kolmogorov-Arnold-Moser or KAM tori). In order to illustrate the method, we consider charged test particles moving in electrostatic waves. The equation which models the dynamics in this case is $\ddot{x} = \sum_{i=1}^{p} a_i k_i \sin(k_i x - \omega_i t + \varphi_i)$ where $a_i$, $k_i$, $\omega_i$ and $\varphi_i$ are respectively the amplitudes, wave numbers, frequencies and phases of the $p$ electrostatic waves. Generically, the dynamics of the particles governed by this equation is a mixture of regular and chaotic behaviours mainly depending on the amplitude of the waves [16]. A Poincaré section of the dynamics (a stroboscopic plot of selected trajectories) is depicted on Fig. 1 for two waves ($p = 2$). A large zone of chaotic behaviour occurs in between the primary resonances (nested regular structures). If one considers a beam of initially monokinetic particles, this zone is associated with a large spread of the velocities after some time (left hand side panel of Fig. 1) since the particles are moving in the chaotic sea created by the overlap of the two resonances [17].

Here the problem of control is to find the right set of electrostatic waves such that the controlled dynamics given by $\ddot{x} = \sum_{i=1}^{p+q} a_i k_i \sin(k_i x - \omega_i t + \varphi_i)$ with $q$ additional waves (called controlling waves) becomes regular. In general, this problem has obvious solutions with additional waves with amplitudes of the same order as the initial ones. However, for energetic purposes, the additional waves must have amplitudes which are much smaller than the initial ones. The idea is to just slightly modify the system (by a specifically designed small control term) so as to drastically change its dynamics from chaotic to regular behaviour. In particular one aims at building barriers to transport that prevent large scale chaos to occur in the system.

Recently a new method of control of hamiltonian chaos has been proposed based on perturbation theory and Lie algebra [18-19]. We consider hamiltonians of the form $H = H_0 + \varepsilon V$, i.e. a regular dynamics described by $H_0$ perturbed by a potential $\varepsilon V$. We consider Hamiltonians in action-angle variables $(I, \varphi)$ written as $H(I, \varphi) = H_0(I) + \varepsilon V(I, \varphi)$ where $\varphi \in (\mathbb{R}/2\pi\mathbb{Z})^L$ and $I \in \mathbb{R}^L$. Recall that for $\varepsilon = 0$, since $H_0$ depends only on the actions $I$, its phase space is foliated by invariant tori. By increasing the amplitude $\varepsilon$ of the perturbation from 0, more and more chaotic structures appear and overlap, leading to large scale chaos. The control theory aims at building a control term $f$ which is much smaller than the perturbation $\varepsilon V$ (for instance with $f$ of order $\varepsilon^2$) such that the controlled hamiltonian $H_c = H_0 + \varepsilon V + f$ is integrable or has a more regular behaviour than the original system. We define the frequency vector $\omega(I) = \partial H_0 / \partial I$ and we consider the Fourier expansion of the potential with respect to the angles $V = \sum_{k \in \mathbb{Z}^L} V_k(I) e^{ik\varphi}$.

We also introduce $V_0$ as the projection of the potential $V$ on its resonant modes,

$$V_0 = \sum_{\substack{k \in \mathbb{Z}^L \\ \text{s.t. } \omega \cdot k = 0}} V_k e^{ik\varphi}, \text{ and } W \text{ computed as } W = \sum_{\substack{k \in \mathbb{Z}^L \\ \text{s.t. } \omega \cdot k \neq 0}} \frac{V_k}{i\omega \cdot k} e^{ik\varphi}.$$

The expression of the control term is given by $f = \sum_{n=1}^{\infty} \frac{\varepsilon^{n+1}}{(n+1)!} L_W^n (nV_0 + V)$, where the Lie operator [20] $L_W$ acts on a function $U$ as the Poisson bracket between $W$ and $U$: $L_W U = \frac{\partial W}{\partial \varphi} \cdot \frac{\partial U}{\partial I} - \frac{\partial U}{\partial \varphi} \cdot \frac{\partial W}{\partial I}$, and $L_W^n = L_W \left( L_W^{n-1} \right)$ denotes the $n$-th iterate of the operator $L_W$. By using the above control term $f$ which is of order $\varepsilon^2$, the controlled hamiltonian $H_0 + \varepsilon V + f$ is conjugate to $H_0 + \varepsilon V_0$ which is integrable for $\varepsilon$ sufficiently small and for $\omega$ non-resonant. More precisely, we have

$$e^{L_W} (H_0 + \varepsilon V + f) = H_0 + \varepsilon V_0.$$

Applied to our particular example, this method of control provides a set of controlling waves which drastically reduce the chaotic zone and the spread of the distribution of the velocities by building barriers in phase space. For a two-wave system, one additional controlling wave given by the first term of the control series is sufficient to drastically reduce the chaotic behaviour of the initial dynamics. This additional wave has an amplitude $a_3 = 2a_1 a_2 (\omega_1 / k_1 - \omega_2 / k_2)^{-2}$, a wave number $k_3 = k_1 + k_2$, a frequency $\omega_3 = \omega_1 + \omega_2$ and a phase $\varphi_3 = \varphi_1 + \varphi_2$. For the case considered in the numerical implementation (right hand side panels of Fig. 1), the amplitude of the controlling wave $a_3$ is less than 10% of the initial amplitudes $a_1$ and $a_2$. Even if the modification of the equations of motion is small, the numerical results (right panels of Fig. 1) show a clear trend to reducing the chaotic zone by the creation of KAM tori acting as barriers in phase space. A beam of particles with constant initial velocity undergoes a very limited spread in velocity compared to the case without control, due to the presence of confining regular structures.

We test experimentally our method of control on a beam of electrons produced by a long Travelling Wave Tube (TWT) [21-22], as depicted in Fig. 2. This device has been extensively used to mimic beam plasma interaction [23-24]. It consists of three main elements: an electron gun, a slow wave structure (SWS) formed by a helix with axially movable antennas, and an electron velocity analyzer. The electron gun produces a quasi-monokinetic electron beam, as shown in Fig. 3a. The electron beam, with radius 3 mm, propagates along the axis of the SWS and is confined by a strong axial magnetic field of 0.05 T. Waves are launched with an antenna at the gun side of the SWS. The SWS consists of a 4 m long wire helix, made of a 0.3 mm diameter Be-Cu wire with a radius of 11.3 mm and a pitch of 0.8 mm, rigidly held together by three threaded alumina rods enclosed in a glass vacuum tube evacuated at both end by two ion pumps at a typical pressure of 2 x $10^{-9}$ Torr. A resistive rf termination serves to reduce reflections at each end of the helix. The glass vacuum jacket is enclosed by an axially slotted cylinder that defines the rf ground and ensures that no other empty wave guide modes than the helix modes can propagate. These modes have electric field components along the axis of the helix and axial phase velocities close to the electron beam velocity (approximately the velocity of light multiplied by the tangent of the helix pitch angle). They can be excited by an antenna moving through a cylinder slot and capacitively coupled to the helix in the frequency range from 5 to 95 MHz. The SWS is long enough to allow non linear processes to develop, such as trapping of the beam in the potential troughs of a single wave. Moreover the electron beam density $n_b$ is chosen weak enough to ensure that the beam induces no wave growth and the beam electrons can be considered as test electrons. Finally the cumulative changes of the electron beam distribution are measured with a trochoidal velocity analyzer at the end of the interaction region. A small fraction (0.5 %) of the electrons passes through a hole in the centre of the front collector, and is slowed down by three retarding electrodes. By operating a selection of electrons through the use of the drift velocity caused by an electric field perpendicular to the magnetic field, the direct measurement of the current collected behind a tiny off-axis hole gives the time averaged beam axial energy distribution with an unprecedented resolution [25].

We apply an oscillating signal at a frequency of 30 MHz on the antenna. This signal generates two waves: a helix mode with a phase velocity equal to $v_\phi = 4.07 \times 10^6$ m/s, a beam mode with a phase velocity equal to the beam velocity $v_b$ (in fact two modes with pulsation $\omega = kv_b \pm \omega_b$ corresponding to the beam plasma mode with pulsation $\omega_b = (n_b e^2 / m\varepsilon_0)^{1/2}$, Doppler shifted by the beam velocity $v_b$, merging in a single mode since $\omega_b << \omega$ in our conditions). Fig. 3b shows the measured velocity distribution of the beam after interacting with these two modes over the length of the TWT. The yellow (resp. blue) band gives the size of the resonant domain determined as the trapping velocity width of the helix (resp. beam) mode $v_\phi \pm 2\sqrt{e\Phi_h / m}$ (resp. $v_b \pm 2\sqrt{e\Phi_b / m}$) where $\Phi_h = 2.33$ V (resp. $\Phi_b = 0.17$ V) is the amplitude of the helix (resp. beam) mode determined both from antennas and beam measurements. These two domains slightly overlap and the break up of invariant KAM tori (or barriers to velocity diffusion) [26] results in a large spread of the initially narrow beam of Fig. 3a over the chaotic region (note the change in scale for the vertical axis); only two small bumps remain which correspond to nested regular regions in phase space as shown in Fig. 1.

We now use an arbitrary waveform generator [24] to launch the same signal at 30 MHz and an additional wave as given by the control theory, with a frequency equal to 60 MHz, an amplitude given by the control theory formula, and a well defined phase with respect to the main signal. The beam velocity of Fig. 3 is also chosen in such a way that the wave number of the helix mode at 60 MHz properly satisfies the wave number relation. As observed on Fig. 3c, where the red arrow indicates the phase velocity $v_c$ of the controlling wave, the beam recovers a large part of its initial kinetic coherence and does not spread in velocity beyond $v_c$ in agreement with theory and as indicated in the numerical simulations of Fig.1.

The control is realized with an additional cost of energy which corresponds to less than 1% of the initial energy of the two-wave system. We stress the importance of a fine tuning of the parameters of the theoretically computed control term (e.g., amplitude, phase velocity) in order to realize the apt modification for a system to operate in a regular regime. It opens the possibility to practically achieve control of a wide range of systems at low additional cost of energy.

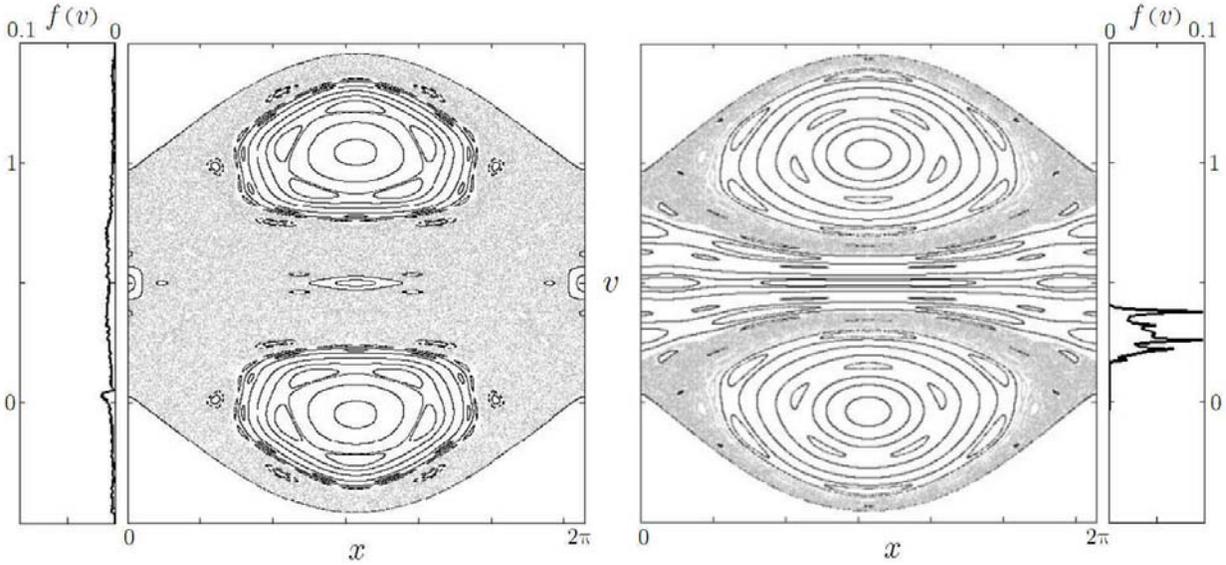

**Figure 1:** Poincaré surface of section and (normalized) probability distribution functions $f(v)$ of the velocity for the dynamics given by $\ddot{x} = \varepsilon\left(\sin x + \sin\left(x - t\right)\right)$ (left panel) and by the controlled dynamics $\ddot{x} = \varepsilon\left(\sin x + \sin\left(x - t\right)\right) + 4\varepsilon^2 \sin(2x - t)$ (right panel) for $\varepsilon = 0.045$. The large chaotic zone between the nested regular structures (left panel) is drastically reduced by the addition of the control. As a consequence, we observe a strong reduction of the kinetic decoherence of the beam.

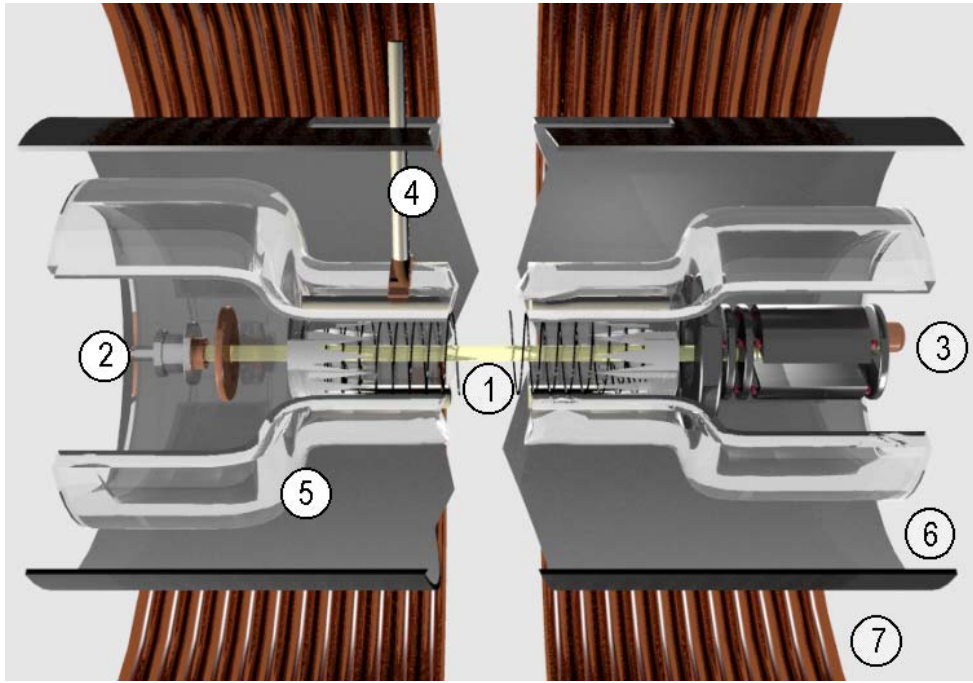

**Figure 2:** Rendering of the Travelling Wave Tube: (1) helix, (2) electron gun, (3) trochoidal analyzer, (4) antenna, (5) glass vacuum tube, (6) slotted rf ground cylinder, and (7) magnetic coil.

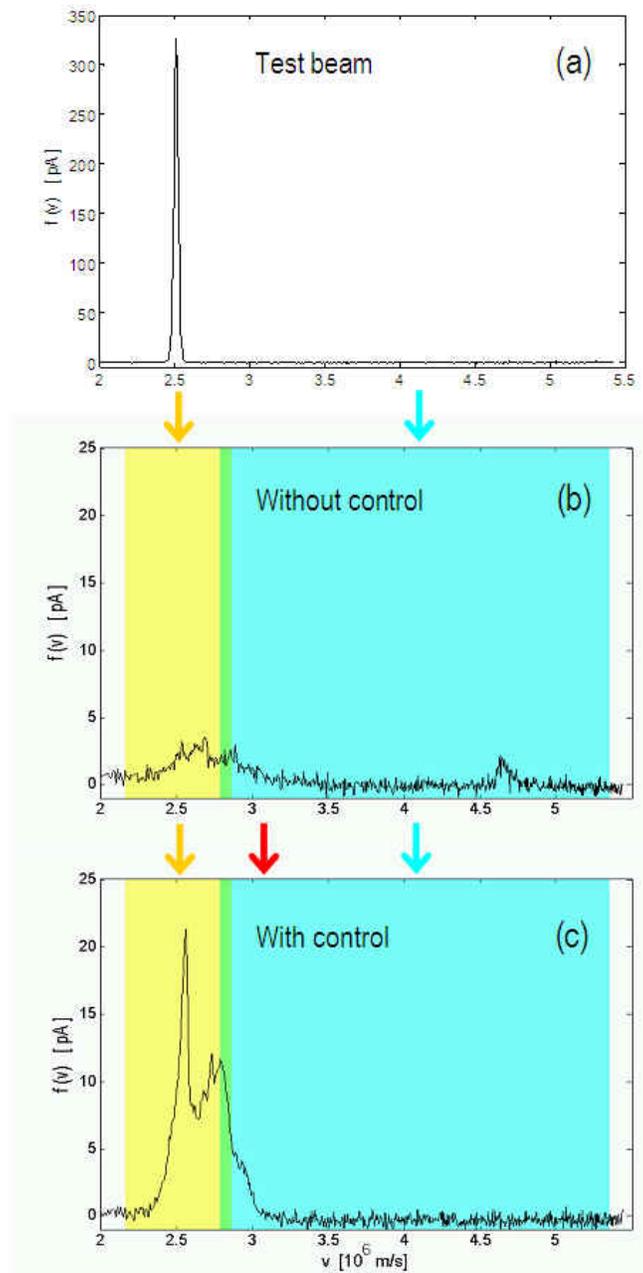

**Figure 3:** Beam velocity distribution function at the output of the TWT: **a**, test beam without electrostatic wave, **b**, with helix mode (blue) and beam mode (yellow) at 30 MHz (phase velocity given by upper arrow and trapping domain of each mode given by coloured bands), **c**, with an additional controlling wave at 60 MHz and phase velocity given by red upper arrow.

**Acknowledgements** Two authors (F. D. and A. M.) are grateful to J-C. Chezeaux, B. Squizzaro and D. Guyomarc'h for their skilful technical assistance.